\documentclass[prl,twocolumn,amsfonts]{revtex4}
\usepackage{amsfonts}
\usepackage{amsmath}
\usepackage{amssymb}
\usepackage{graphics,graphicx}
\usepackage{psfrag}
\usepackage{epsf}

\def\prl{ Phys. Rev. Lett.}

\def\etal{{\it et al.\/}}

\newcommand{\be}{\begin{equation}}
\newcommand{\ee}{\end{equation}}
\newcommand{\bea}{\begin{eqnarray}}
\newcommand{\eea}{\end{eqnarray}}
\newcommand{\bs}{\begin{equation}\begin{split}}
\newcommand{\es}{\end{split}\end{equation}}

\begin{document}

\author{H. Landa$^1$, S. Marcovitch$^1$, A. Retzker$^2$,
M. B. Plenio$^2$ and B. Reznik$^1$}
\title{Quantum Coherence of Discrete Kink Solitons in Ion Traps}
\affiliation{$^1$School of Physics and Astronomy,
Raymond and Beverly Sackler Faculty of Exact Sciences,
Tel-Aviv University, Tel-Aviv 69978, Israel\\
$^2$Institute for Mathematical Sciences, Imperial College London,
London SW7 2PG, United Kingdom, \\and QOLS, The Blackett Laboratory,
Imperial College London, London SW7 2BW, United Kingdom, \\ and Institut f{\"u}r Theoretische Physik, Universit{\"a}t Ulm, D-89069 Ulm.}

\begin{abstract}

We propose to realize quantized discrete kinks with cold trapped ions. We show that long-lived solitonlike configurations are manifested as deformations of the zigzag structure
in the linear Paul trap, and are topologically protected in a circular trap with an odd number of ions. We study the quantum-mechanical time evolution of a high-frequency, gap separated internal mode of a static kink and find long coherence times when the system is cooled to the Doppler limit. The spectral properties of the internal modes make them ideally suited for manipulation using current technology. This suggests that ion traps can be used to test quantum-mechanical effects with solitons and explore ideas for the utilization of the solitonic internal-modes as carriers of quantum information.

\end{abstract}
\maketitle


Solitons are localized configurations of nonlinear systems which are nonperturbative and topologically protected \cite{Rajaraman}. Quantum-mechanical properties of solitons, such as squeezing, have been predicted and measured in optical systems \cite{shelby_exp}. Quantum dynamics has been observed with a single Josephson junction soliton \cite{Vortex}. In waveguide arrays \cite{silberberg,segev} and Bose-Einstein condensates \cite{latticeBEC} solitons are mean field solutions, localized to a few sites of a periodic potential. In chains of coupled particles, solitons are discrete spatial configurations, as in the Frenkel-Kontorova (FK) model \cite{FK,BraunKivsharBook}.


Discrete solitons of the FK model and its generalizations
are referred to as kinks. An important property of kinks is the
existence of localized modes. One mode is the kink's translational
`zero-mode', whose frequency generally rises above zero.
Other localized modes are known as `internal modes' \cite{kivshar2,Takeno}.
Physically they describe `shape-change' excitations of the kink and
typically they are separated by an energy gap
from other long-wavelength phononic modes. It was suggested to use the internal mode as a carrier of quantum information \cite{Markovitz}.

\begin{figure}[ht]
\center {\includegraphics[width=3.4in]{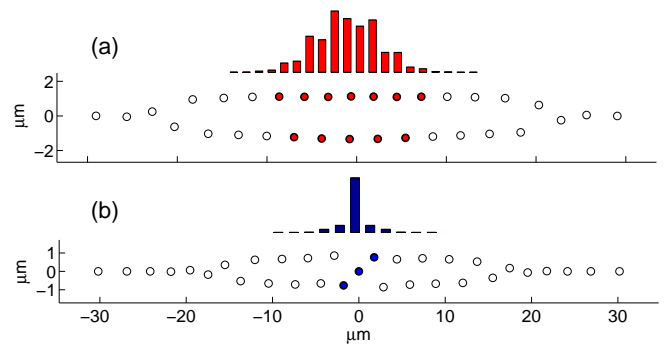}
\caption
{Metastable linear trap configurations with 33 ions. (a) An extended
kink. The localized internal mode (red bars) involves $\sim{10}$ ions.
(b) A highly discrete 'odd' kink. A localized internal oscillation
(blue bars) involves $\sim3$ ions.}
\label{KinkConfigs}}
\end{figure}

Quantum information processing in ion traps \cite{review} has dramatically
improved over the last decades \cite{Leibfried2005,Hartmut6}.
Recently there has been a considerable interest in using
trapped ions for quantum simulation of various systems such as
spin-chains \cite{wunderlic1,Porras1,Tobias} and Bose-Hubbard models \cite{Porras2},
field models \cite{Retzker2005}, cosmological effects \cite{Alsing} and
black holes \cite{birger}. It was suggested to realize a 1D generalization of the FK model by adding an external periodic potential to an ion trap \cite{EuroPhys}.

In this Letter we demonstrate that quantum coherence in static discrete kinks can be observed with ordinary
Paul traps without external additions. We explore quasi-2D
discrete kinks resembling those of the zigzag model
\cite{zigzagkinks}. In the linear trap we find
local metastable deformations of the zigzag
structure \cite{Dubin}, as depicted in Fig. \ref{KinkConfigs}, which are long-lived already with a moderate number of ions, $N\gtrsim20$. In a circular trap with an odd number
of ions, similar configurations form the ground state. We study the robustness of a
high-frequency internal mode of the kink against decoherence
in the thermal environment of all the other modes.
With all nonlinear interactions accounted for, we numerically integrate a
non-Markovian master equation, which leads us to our main result: already at the standard
Doppler cooling limit coherence persists in the internal mode
for many oscillations. This could allow a first direct measurement of decoherence time of solitons.



Let us consider $N$ ions trapped either in a linear trap or
in a circular ring trap. Throughout this Letter we use
nondimensional units by employing the natural length scale
$d =\left( e^{2} /m \nu^{2} \right)^{1/3}$ and time scale
$1/\nu$, where $m$ is the ions' mass and $\nu$ is the axial
(radial) trapping frequency in the linear (circular) trap.
In the linear case the trapping potential can be expressed
as $V= \sum_{i}^{N}\frac{1}{2} \left(x_i^2 + \beta y_i^2\right)$.
At sufficiently high transverse trapping $\beta$, the ions
crystalize in a one-dimensional chain along the $x$-axis.
When $\beta$ is lowered, the ions undergo
a second order phase transition \cite{Morigi2} and the lowest
energy state has a zigzag shape with interesting quantum
mechanical properties \cite{Retzker2008}. However, depending
on $\beta$, other metastable configurations exist. We have
identified two additional types of configuration that are
not destroyed by thermal fluctuations at up to $\sim15$ times
the Doppler temperature. Figure \ref{KinkConfigs}(a) shows
an extended soliton-like deformation ($\beta=40$), with a
cross-over of the upper ions of the zigzag above the lower
ones. In Fig \ref{KinkConfigs}(b) one ion is forced out
of the zigzag to form a dense defect at the center ($\beta=85$).

\begin{figure}[ht]
\center{ \includegraphics[width=3.1in]{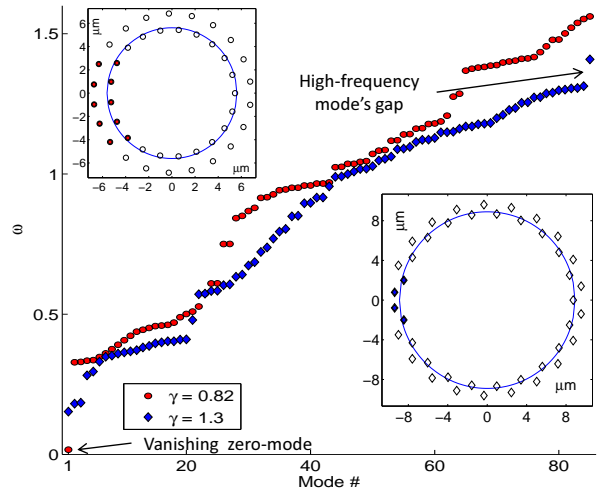}
\caption
{Dispersion-relations in the circular trap. At $\gamma=0.82$ (red circles and upper inset) the kink is extended, there is no gap at the top of the spectrum and the translational mode's frequency approaches zero. At $\gamma=1.3$ (blue diamonds and lower inset) the kink is localized on 4 ions and the localized modes at the top and bottom of the spectrum are gap-separated. The mode corresponding to rotational symmetry of the entire trap has been omitted.}
\label{Dispersion}
}
\end{figure}

In the ring trap, ions are confined at uniform density
around a circle and the radial confining potential is
\be V= \sum _{i}^{N}{\frac{1}{2} \left(\left\| \vec{R}_{i}
\right\| -\gamma\frac{N}{2\pi} \right)^{2}},\label{RingPotential}\ee
where $\vec{R}_{i}\equiv \left(x_i,y_i\right)$ and $\gamma$
measures the strength of the radial trapping, independently
of the number of ions. $\gamma\propto \nu^{2/3}$ is varied by changing the radial trapping frequency. The lowest energy configuration with an odd number of ions can be one of a few types. At high $\gamma$, close to the phase-transition from a 1D chain ($\gamma\approx1.6$), the kink is localized with two ions facing {\it inside }the zigzag shape. At lower values of $\gamma$ the two kink-core ions lie {\it outside } of the zigzag, as shown in Fig. \ref{Dispersion}, lower inset. When lowering $\gamma$ further, an extended kink is formed (Fig. \ref{Dispersion}, upper inset). The ring topology protects kinks in this trap from breaking.

\begin{figure}[ht]
\center{ \includegraphics[width=3.2in]{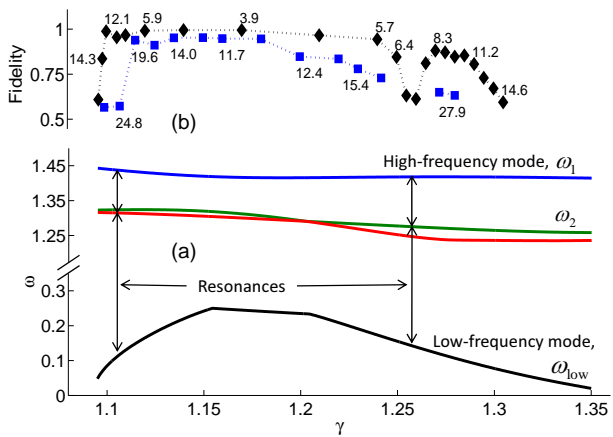}
\caption
{Mode frequencies and fidelity analysis. (a) The frequencies of the lowest and the three highest modes as a function of the trap parameter in the 'odd' kink configuration (see text for details). (b) The lowest value of fidelity obtained for simulations of 50 oscillations of the high-frequency mode, at different parameter values. Black diamonds: at inter-ion separation of $1.5\mu{m}$. Blue squares: at twice that distance. The numbers indicate the population of thermal phonons in the low-frequency mode.
\label{FidelityAnalysis}}
}
\end{figure}

In the absence of a kink the linearized frequency spectrum
of the normal-modes consists of a phonon band terminated at
a cutoff frequency that depends on the ion density. The presence
of the kink causes a few highly localized modes to split
away from the rest of the spectrum. In a large range of values for the trap
parameter, there exists a localized mode lying above the band
and separated by a gap (Figs. \ref{Dispersion}, \ref{FidelityAnalysis}(a)). This mode corresponds to
out-of-phase oscillations of the ions in the kink's core.
It is this "high-frequency mode" which we suggest for coherent manipulations.

Expanding the Hamiltonian in a perturbative series about the classical
kink configuration, $\vec{R}_{i} = \vec{R}^{0}_{i} + \delta\vec{R}_{i}$,
we switch variables to the normal coordinates which diagonalize the
harmonic part. Keeping nonlinear terms up to the fourth-order we
proceed with canonical quantization of the normal coordinates to get
\bea
H=H_{free} +\frac{1}{3!} \sum_{ijk}L_{ijk} \hbar ^{\frac{3}{2}} {\left(8\omega_{i} \omega_{j} \omega_{k}\right)}^{-\frac{1}{2}} \Theta_i\Theta_j\Theta_k & \nonumber \\
+\frac{1}{4!} \sum _{ijkl}M_{ijkl} \hbar ^{2} \left(16\omega_{i} \omega_{j} \omega_{k} \omega_{l} \right)^{-\frac{1}{2}} \Theta_i\Theta_j\Theta_k\Theta_l, \label{H}
\eea
where $H_{free}$ is the Hamiltonian of free-phonons, $\Theta_j\equiv (a_{j}^\dag + a_{j})$, $a_{j} ^{\dag }$ is the creation operator of the normal mode $j$ with frequency $\omega_j$ (a linear combination of all the $\delta\vec{R}_{i}$), and $\hbar\sim 2\times10^{-5}$ is nondimensional.

We now briefly outline the key steps of a derivation of a
master equation modelling the coherent quantum-mechanical time
evolution of the mode of interest. We follow
the notation in \cite{Carmichael}. We divide the modes into the
'system' -- consisting of the high-frequency mode ($\omega_1$), and the
'bath' -- all other modes, and split the Hamiltonian $H=H_{S} +H_{B} +H_{SB} $
into 3 parts: for the system, bath and system-bath interaction,
respectively.
We next apply the Born approximation in the Liouville-von Neumann equation,
assuming the factorization
$\tilde{\rho }\left(0\right)\otimes\chi _{B}$
where $\tilde O$ denotes an operator in the interaction picture and $\chi _{B}$ is a thermal density matrix for the bath, yielding
\bea
\lefteqn{ \dot{\tilde{\rho }}\left(t\right)=}\nonumber\\ &-\frac{1}{\hbar ^{2} } tr_{B} \left\{\left[\tilde{H}_{SB} \left(t\right),\int _{0}^{t}dt'\left[\tilde{H}_{SB} \left(t'\right),\tilde{\rho }\left(t'\right) \otimes \chi _{B} \right] \right]\right\}.\label{eq:master1}
\eea
Expressing the interaction as a sum of terms , each consisting of a system operator $s_{\alpha}$ multpilying a bath operator $B_{\alpha}$, we have
$\label{eq:HSB}\tilde{H}_{SB}(t)=\hbar \sum _{\alpha }\tilde{s}_{\alpha }(t)\tilde{B}_{\alpha }(t)$. Assuming the bath to remain in thermal equilibrium we take a free-evolution Hamiltonian for $H_B$. In $H_S$, however,  we keep all terms. In $H_{SB}$, we include operators up to quadratic order from both the system and the bath. Thus, $s_{\alpha } \in \{\Theta_1,\Theta_1^{2}\}$, and correspondingly,
\bea \tilde{B}_{1} (t)&= \frac{1}{2} \sum _{jk\ne 1}L_{1jk} \hbar^{\frac{1}{2}}\left(8\omega _{1} \omega _{j} \omega _{k}  \right)^{-\frac{1}{2}} \tilde{\Theta}_j \tilde{\Theta}_k\nonumber \eea\bea \tilde{B}_{11} (t) & = & \frac{1}{2} \sum _{k\ne 1}L_{11k} \hbar^{\frac{1}{2}}\left(8\omega _{1}^2 \omega _{k} \right)^{-\frac{1}{2}} \tilde{\Theta}_k \nonumber \\ & +& \frac{1}{4} \sum _{kl\ne 1}M_{11kl} \hbar \left(16\omega _{1}^2 \omega _{k} \omega _{l} \right)^{-\frac{1}{2}} \tilde{\Theta}_k \tilde{\Theta}_l. \nonumber \eea

\begin{figure}[ht]
\center{ \includegraphics[width=3.2in]{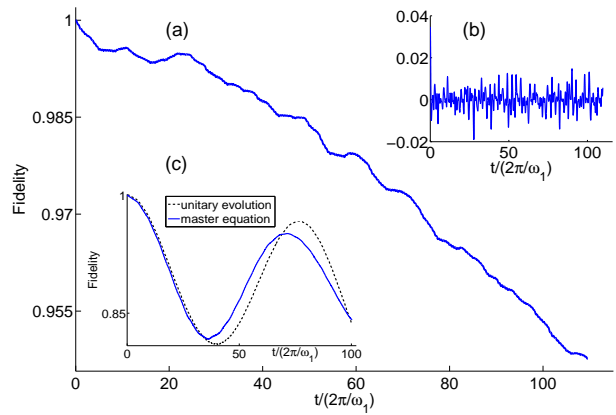}
\caption
{(a) Fidelity of coherent oscillations of the high-frequency mode in the linear trap ($\beta=85$). (b) $Re\langle \tilde{C}_{11}(\tau)\rangle$ of the same configuration. (c) Fidelity near a resonant configuration, using the master equation (blue), and a unitary evolution (dashed).}
\label{Fidelity}
}
\end{figure}

We define the renormalized bath correlation functions \bea\tilde{C}_{\alpha \beta } \left(t-t'\right)=\left\langle\left(\tilde{B}_{\alpha } (t)-\langle\tilde{B}_{\alpha}\rangle\right) \left(\tilde{B}_{\beta}(t')-\langle\tilde{B}_{\beta}\rangle\right)\right\rangle_{B}\label{eq:Cab}\eea
which do not decay [Fig. \ref{Fidelity}(b)]. This is due to the discreteness and cutoff of the bath spectrum, in addition to the nonlinearity of the interaction. We therefore proceed with a non-Markovian treatment.
This results in the integro-differential equation
\bea
\dot{\tilde{\rho }}=-\sum _{\alpha ,\beta }\left[\tilde{s}_{\alpha } \tilde{S}_{\beta \alpha } -\tilde{S}_{\beta \alpha } \tilde{s}_{\alpha } +\tilde{S}^{\dag } _{\beta \alpha } \tilde{s}_{\alpha } -\tilde{s}_{\alpha } \tilde{S}^{\dag } _{\beta \alpha } \right]\label{eq:master2}
\eea
where $\tilde{S}_{\alpha \beta } \left(t\right)\equiv \int _{0}^{t}\tilde{s}_{\alpha } \left(t'\right)\tilde{\rho}\left(t'\right)\tilde{C}_{\alpha \beta } \left(t-t'\right)dt'$.

We solve eq. (\ref{eq:master2}) numerically, taking 33 $Ca^{+}$ ions in the configuration of figure \ref{KinkConfigs}(b).
With $\nu_x/2\pi=0.88 MHz$ and $\nu_y/2\pi=8.1 MHz$, the mode frequencies are $\omega_1/2\pi=11.5 MHz$ for the high-frequency mode, $\omega_2/2\pi=10.6 MHz$ for the next mode, and $\omega_{\rm low}/2\pi=2.1 MHz$.
The inter-ion separation at the kink center is $1.7\mu{m}$. We 
set the temperature of the bath to the Doppler cooling limit
$T_{\rm Doppler} =2\pi\times10 MHz$, and the low-frequency mode has 4.3 phonons
\cite{heating}. In $\rho\left(0\right)$ we assign a representative
superposition state $\left|0\right\rangle + \left|1\right\rangle$.
This state can be created after sideband cooling of the internal
mode and initialization using quantum information techniques \cite{Leibfried2005,Hartmut6}. This must be done slowly compared to the inverse of the energy gap separation of this mode ($2\pi\times0.9MHz$, or $\sim \omega_1/12$ in the above example). In Fig. \ref{Fidelity}(a)
we show the fidelity \cite{fidelity} of the system's evolution
versus time. The fidelity is calculated \cite{qlib} with reference
to an isolated free phonon, for which there is an oscillation of the relative phase between the levels $\left|0\right\rangle$ and $ \left|1\right\rangle$.
The fidelity remains very high for the simulated $\sim 100$ periods
of oscillation. This would allow the state initialization to be performed with a high fidelity.

As one test of our results, we compare the master
equation approach to a direct unitary calculation involving only
the three modes which are expected to dominate the nonlinear
process near a resonance as in figure \ref{FidelityAnalysis}.
The agreement depicted in figure \ref{Fidelity}(c) indicates
that the master equation indeed captures the evolution very
accurately, and allows one to simulate the coherent affect
of the thermal bath modes on the internal-mode. Far from the resonances, the Born approximation is valid as long as $\epsilon \equiv \Delta E_1/E_{\rm bath} \ll 1$, where $\Delta E_1$ is the energy leaking from the system into the bath, and $E_{\rm bath}$ can be estimated -- for the worst case -- using the low-frequency mode. In the simulation of figure \ref{Fidelity}(a), $\epsilon$ is only a few percent, so the Born approximation is justified.

In order to analyze the dependence of the coherence on the trap
parameter, we take the limit of a large circular trap with vanishing
curvature, which is also the limit of a large linear trap with fixed
ion density in the center. This is achieved using `periodic'
boundary conditions such that the longitudinal distance between any two ions is evaluated modulo half the chain length. The soliton configuration
analyzed is of the same type as in Fig. \ref{KinkConfigs}(b).
Figure \ref{FidelityAnalysis}(b) shows the fidelity as a function
of $\gamma$ [which is independent of the number
of ions, see Eq. \ref{RingPotential}]. The fidelity remains high
provided that $\omega_{\rm low}$ is not too low [Fig. \ref{FidelityAnalysis}(a)],
and that there is no strong resonance, $\omega_1 \approx \omega_j
+ \omega_{\rm low}$, where the largest interaction coefficient is
with $j=2$, a partly-localized phonon. The resonance on the left
is seen to have a much weaker effect, owing to a smaller matrix-element. The loss
of fidelity at the resonances is mainly due to energy relaxation
from the high-frequency mode, while away from resonance the
off-diagonal elements grow considerably, expressing the loss of
phase coherence.

We continue with a discussion of various aspects of our
treatment. The geometric coefficients $M_{ijkl}$ and
successive terms omitted from the series expansion in
eq. (\ref{H}), are not necessarily decreasing compared
with the third order $L_{ijk}$. However, powers of
$\sqrt{\hbar} \sim 1/230$ dominate the convergence at
low excitations, so already the contribution of
quartic terms is negligible in general. The Hilbert space
of the system is truncated at dimensions $8-10$, which we
have checked to have negligible effect on the results
because of the low phonon-numbers.

The dominant contribution to decoherence comes from multi-phonon
processes involving the localized high and low modes. Powers of
${\omega_{\rm low}}^{-1/2}$ enter the coefficients of such interactions
which grow stronger as $\omega_{\rm low}\to 0$. Since $\omega_1$ depends
only weakly on the trap parameter (at a given kink type), while
$\omega_{\rm low}$ and the gap are tunable, the highest fidelity is
achieved at the center area of figure \ref{FidelityAnalysis},
where the gap is large and $\omega_{\rm low}$ is high and far from
the strong resonances. These considerations hold when increasing $N$
at a given geometry, as the localized modes do not change their
frequency or strength of local interactions. The number of all
resonant processes grows proportionally to the phonon density
($\sim N$), but the strength of each one drops as $\sim 1/\sqrt{N}$
for every plane-wave phonon involved. This leaves only the
weakly coupled third-order resonances, near $\omega_j \approx \omega_1 - \omega_{\rm low}$,
with a contribution scaling like $N/\sqrt{N}=\sqrt{N}$. In addition, there
are the {\it off-resonant} couplings with phonons whose frequencies
approach zero. Still, these two contributions will pose no problem
up to at least a couple of hundreds of ions.

Production of kink configurations in the
linear trap can be achieved by a fast temporal variation of
the transverse potential. Numerical simulations indicate that
this process together with simultaneous cooling indeed leads
to creation of kinks [Fig. \ref{KinkConfigs}(b)], which remain stable at
temperatures below $\sim 15T_{\rm Doppler} $.

Finally, our results should hold in other discrete nonlinear models. In particular, long coherence times have been obtained for the FK and discrete $\phi^4$ models \cite{thesis}.

To conclude, we have demonstrated that stable classical
kink configurations exist in linear traps with $\gtrsim20$ ions,
as well as in circular traps. Our results suggest that coherence of solitonic internal modes is preserved for surprisingly
long durations in Doppler cooled traps. The unique scale independent
properties of the internal modes, specifically their gap separation
from the phonon band, their localization to a few ions and their
high frequency, suggest that such coherences can be measured and
manipulated using existing ion trap techniques. This indicates that trapped-ion solitons may be useful for generating entanglement \cite{Markovitz,MalomedLewenstein,LaiLee}, and implementing quantum information processing
in large systems.

B.R. acknowledges the support by Israeli Science Foundation
grants 784/06, 920/09 and the Israeli German Foundation grant I-857. A.R. acknowledges the support of EPSRC project number EP/E045049/1. M.B.P. acknowledges support by the EU Integrated Project QAP, the Royal Society and an Alexander-von-Humboldt foundation.



\begin{thebibliography}{34}
\expandafter\ifx\csname
natexlab\endcsname\relax\def\natexlab#1{#1}\fi
\expandafter\ifx\csname bibnamefont\endcsname\relax
  \def\bibnamefont#1{#1}\fi
\expandafter\ifx\csname bibfnamefont\endcsname\relax
  \def\bibfnamefont#1{#1}\fi
\expandafter\ifx\csname citenamefont\endcsname\relax
  \def\citenamefont#1{#1}\fi
\expandafter\ifx\csname url\endcsname\relax
  \def\url#1{\texttt{#1}}\fi
\expandafter\ifx\csname urlprefix\endcsname\relax\def\urlprefix{URL
}\fi \providecommand{\bibinfo}[2]{#2}
\providecommand{\eprint}[2][]{\url{#2}}

\bibitem{Rajaraman} R. Rajaraman, {\it Solitons and Instantons},
North Holland, Elsevier Science Publisher, Amsterdam (1982).

\bibitem{shelby_exp} S. J. Carter, \etal, Phys. Rev. Lett., {\bf58}, 1841 (1987). M. Rosenbluh and R. M. Shelby, Phys. Rev. Lett. {\bf 66}, 153 (1991).

\bibitem{Vortex} A. Wallraff \etal, Nature {\bf 425}, 155 (2003).

\bibitem{silberberg} H. S. Eisenberg \etal, Phys. Rev. Lett. {\bf 81},
3383 (1998).

\bibitem{segev} J. W. Fleischer \etal, Nature {\bf422}, 147 (2003).

\bibitem{latticeBEC} A. Trombettoni and A. Smerzi, Phys. Rev. Lett. 86, 2353 (2001)

\bibitem{FK} Ya. Frenkel and T. Kontorova, Phys. Z. Sowjetunion {\bf{13}}, 1 (1938).

\bibitem{BraunKivsharBook} O. M. Braun and Y. S. Kivshar, {\it The Frenkel-Kontorova Model}, Springer (2004).

\bibitem{kivshar2} O.M. Braun, Y.S. Kivshar and M. Peyrard, Phys. Rev. E {\bf56}, 6050 (1997).

\bibitem{Takeno} A. J. Sievers and S. Takeno, Phys. Rev. Lett. 61, 970 (1988)

\bibitem{Markovitz} S. Marcovitch and B. Reznik, Phys. Rev. A {\bf 78}, 052303 (2008).

\bibitem{review} For a recent review see, e.g., M. \v{S}a\v{s}ura and V. Bu\v{z}ek, J. Mod Opt. {\bf 49}, 1593 (2002).

\bibitem{Leibfried2005} D. Leibfried \etal, Nature {\bf 438}, 639 (2005).
\bibitem{Hartmut6} H. H\"{a}ffner \etal, Nature {\bf 438}, 643 (2005).

\bibitem{wunderlic1} F. Mintert and C. Wunderlich, \prl {\bf 87} 257904 (2001).

\bibitem{Porras1} D. Porras and J.I. Cirac, Phys. Rev. Lett. {\bf 92}, 207901 (2004).

\bibitem{Tobias} A. Friedenauer \etal, Nature Physics 4, 757 - 761 (2008).

\bibitem{Porras2} D. Porras and J.I. Cirac, Phys. Rev. Lett. {\bf 93}, 263602 (2004).

\bibitem{Retzker2005} A. Retzker, J.I. Cirac and B. Reznik, Phys. Rev. Lett. {\bf 94}, 050504 (2005). L. Lamata \etal,
     Phys. Rev. Lett. {\bf 98}, 253005 (2007).
\bibitem{Alsing} P. M. Alsing, J. P. Dowling, and G. J. Milburn. Phys. Rev. Lett. {\bf 94}, 220401 (2005). R. Sch\"utzhold \etal, Phys. Rev. Lett. {\bf 99}, 201301 (2007).


\bibitem{birger} B. Horstmann \etal, arXiv:0904.4801.

\bibitem{EuroPhys} I. Garc\'{\i}a-Mata et al., Eur. Phys. J. D {\bf 41}, 325 (2007)

\bibitem{zigzagkinks} Oleg M. Braun,  Yuri S. Kivshar, Phys. Rev. B {\bf 44}, 7694 (1991).
O. M. Braun \etal, Phys. Rev. B {\bf48}, 3734 (1993).

\bibitem{Dubin} D. H. E. Dubin and T. M. O'Neil, Rev. Mod. Phys. {\bf 71},
87 (1999).

\bibitem{Morigi2} S. Fishman \etal \prb {\bf 77}, 064111 (2008).

\bibitem{Retzker2008} A. Retzker \etal, \prl {\bf 101},  260504 (2008).

\bibitem{Carmichael} H. Carmichael, {\it Statistical Methods in Quantum Optics}, Springer-Verlag, Berlin (1999).

\bibitem{heating} The small transverse departures of the ions are not expected to cause significant micromotion heating : B. E. King \etal, Phys. Rev. Lett. {\bf 81}, 1525 (1998).

\bibitem{fidelity} For density operators $\rho$ and $\chi$, $F\equiv tr\sqrt{\rho^{1/2}\chi\rho^{1/2}}$.

\bibitem{qlib} S. Machnes, \emph{QLib}, arXiv:quant-ph/0708.0478.

\bibitem{thesis} H. Landa, Master's thesis, Tel Aviv University, arXiv:0910.0109.

\bibitem{LaiLee} R. K. Lee, Y. Lai and B. A. Malomed, Phys. Rev. A  {\bf 71}, 013816 (2005).

\bibitem{MalomedLewenstein}  M. Lewenstein and B. A. Malomed, arXiv:0901.2836 [New J. Phys. (to be published)].










%


\end{thebibliography}
\end{document}